\documentclass[aps,prb,amsfonts,showpacs,floats,floatfix,twocolumn,nofootinbib,byrevtex]{revtex4}

\usepackage{epsfig}
\usepackage{amsmath}
\usepackage{amsfonts}
\usepackage{color}
\usepackage{ulem}
\usepackage{graphicx,slashbox}

\usepackage{ulem}
\newcommand{\MDcomm}[1]{{\color{black} #1}}

\newcommand{\PGRcomm}[1]{{\color{black} #1}}

\begin{document}

\title{Time-dependent density-functional theory with 
self-interaction correction}
\author{J. Messud$^1$, P. M. Dinh$^1$, P.-G. Reinhard$^2$, and E. Suraud$^1$}
\affiliation{
  1) Laboratoire de Physique Th\'eorique, IRSAMC, CNRS, Universit{\'e} de
  Toulouse,  F-31062 Toulouse, France \\
  2) Institut f\"ur Theoretische 
  Physik,
  Universit{\"a}t Erlangen, D-91058 Erlangen,
  Germany 
}

\date{\today} 
\begin{abstract}
We discuss an extension of time-dependent density-functional theory by
a self-interaction correction (SIC).  A strictly variational
formulation is given taking care of the necessary constraints.  A
manageable and transparent propagation scheme using two sets of
wavefunctions is proposed and applied
to laser excitation with subsequent ionization of a dimer molecule.
\end{abstract}

\maketitle 

Density Functional Theory (DFT)~\cite{Hoh64,Par89,Dre90,Koh99} has
evolved over the last decades to a standard theoretical tool for the
description of electronic properties in many physical and chemical
systems, especially in systems with sizable numbers of electrons.  The
extension of DFT to  Time-Dependent situations
(TDDFT)~\cite{Run84,Gro90,Mar04} 
is a more recent achievement still  motivating many 
investigations, both for formal and practical aspects
\cite{Mar06}. It turns out that TDDFT is one of the few, well founded
theories, allowing to describe dynamical scenarios in complex systems, 
{which is a key issue for understanding microscopic mechanisms, beyond 
mere energetic considerations.}
This is especially true if electron emission comes into play as, e.g.,
in case of {irradiation processes.}

A practical treatment of DFT, and even more so TDDFT, requires simple
approximations to the exchange and correlation functional. The
simplest one is the Local Density Approximation (LDA) which proved
very useful in calculations of structure and low-amplitude excitations
(optical response, direct one-photon processes) \cite{Koh99}.  It is
also often used as a first order approach for more violent dynamical
processes where huge energy deposits lead to a large number of emitted
electrons as, e.g., in clusters subject to intense laser fields or
collided by highly charged particles \cite{Rei03a}. However, LDA is
plagued by a self-interaction error because its Kohn-Sham field
involves the total density including the particle on which the field
just acts.
As a consequence, the LDA produces the wrong Coulomb asymptotic of
the mean field and thus underestimates the ionization potential (IP)
of a system. 
\PGRcomm{This spoils, e.g., the dynamical description  of
excitations involving ionization processes, in particular close to 
the threshold. A correct treatment
requires}  a self-interaction correction (SIC).  Such a SIC
complementing LDA for static calculations was proposed in \cite{Per81}
and has been used since then at various levels of refinement for structure
calculations in atomic, molecular, cluster and solid state physics,
see e.g. \cite{Ped84,Goe97,Polo,Vydrov}. 

The original SIC scheme leads to
an orbital dependent mean field which causes several formal and
technical difficulties. There are attempts to circumvent these
problems by treating SIC with optimized effective potentials (OEP),
see \cite{Kri92,Kue07}. That, however, overrules some crucial physical
features of SIC, particularly the trend to produce localized
single-particle states \cite{Kue07}.

Application of SIC in time-dependent situations are mostly done in
approximate manner, \PGRcomm{linearized \cite{Pac92}, using
averaged-density SIC \cite{Leg02a}, or relying on various versions of
time dependent OEP-KLI \cite{Ull95a,Ton97,Ton01}. The
TDOEP-KLI, however, also suffers from inconsistencies as it leads,
e.g.  to violation of zero force theorem and energy
conservation \cite{Mun07a}}.  The aim of this letter is to present a
thorough variational formulation of \PGRcomm{fully fledged} TDSIC
\PGRcomm{without further approximation}, together with a \PGRcomm{manageable}
propagation scheme which allows to obey all boundary conditions,
\PGRcomm{namely the zero-force theorem, conservation of energy and
orthonormality of the occupied single-particle orbitals.}  {This will
in particular serve as a benchmark for the development and validation
of further approximate treatments for example in the spirit of (TD)OEP
approaches}.  A first application to a one-dimensional molecule proves
the feasibility and stability of the scheme. A comparison with exact
exchange confirms the need \PGRcomm{and satisfying performance of full TDSIC
in ionization dynamics.}


We work in the Kohn-Sham scheme of DFT, 
{built on a set of single particles}
wavefunctions  $\{\psi_\alpha, \, \alpha=1,\ldots,N\}$. 
{In the SIC case, the  starting energy functional reads}
\begin{eqnarray}
  E_\mathrm{SIC}
  =
  E_\mathrm{kin}
  \!+\!
  E_\mathrm{ion}
  \!+\!
  E_\mathrm{LDA}[\rho]
  \!-\!
  \sum_{\beta=1}^{N} E_\mathrm{LDA}[\rho_{\beta}]
\label{eq:fsicen}
\end{eqnarray}
where the electronic kinetic energy $E_\mathrm{kin}$ is complemented by the 
external ionic contribution $E_\mathrm{ion}$ and the LDA approximation 
$E_\mathrm{LDA}[\rho]$ to the exact electron-electron interaction energy.
The densities are further defined as $\rho_\alpha=|\psi_{\alpha}|^2$
and $\rho=\sum_{\alpha} \rho_{\alpha}$.
Note that all summations run over occupied states only. 

The TDSIC \PGRcomm{equations are} obtained from the principle of
stationary action using the SIC energy functional (\ref{eq:fsicen})
\begin{eqnarray}
 0
 &=&
\delta \int_{t_0}^{t}dt' \Big(
 \sum_{\alpha}(\psi_\alpha|\mathrm{i}\hbar \partial_t|\psi_{\alpha})
  -
  E_\mathrm{SIC}
\nonumber\\
  &&
  \qquad\quad\quad
  +
  \sum_{\beta,\gamma}^{}(\psi_{\beta}|\psi_{\gamma})
  \lambda_{\gamma\beta}
 \Big)
 \quad,
\label{eq:varprinconstr}
\end{eqnarray}
within explicitely including the orthonormality constraint with Lagrange 
multipliers $\lambda_{\gamma\beta}$.
{Variation of
$E_\mathrm{SIC}$ with respect to $\psi^*_{\alpha}$ 
leads to single particle equations for the $\psi_\alpha$'s
in which the} 
one-body  Hamiltonian is obtained as
\begin{subequations}
\label{eq:mfham}
\begin{eqnarray}
  \frac{\delta E_\mathrm{SIC}}{\delta\psi_\alpha^*}
  &=&
  \hat{h}_\alpha\psi_\alpha
  \;,\;
  \hat{h}_\alpha
   =
  \hat{h}_\mathrm{LDA}-U_\alpha
  \;,
\\
  \hat{h}_\mathrm{LDA}
  &=&
  -\frac{\hbar^2 \Delta}{2m}
  + U_\mathrm{LDA}[\rho]
  \quad,
\\
  U_\alpha 
  &=&  
  U_\mathrm{LDA}[|\psi_\alpha|^2]
  \quad,
\label{eq:usic}
\\
  U_\mathrm{LDA}[\varrho]
  &=&
  \left.\frac{\delta E_\mathrm{LDA}}{\delta \rho}\right|_{\rho=\varrho}
  \quad.
\end{eqnarray}
The emerging one-body Hamiltonian $\hat{h}_\alpha$ depends on the
state $\psi_\alpha$ on which it acts through the SIC term $U_\alpha$.
The $\hat{h}_\alpha$'s  
can be {simply} recast in a SIC Hamiltonian 
${\hat h}_{\rm SIC}$ by employing projectors as
\begin{eqnarray}
  \hat{h}_\mathrm{SIC}
  &=&
  \hat{h}_\mathrm{LDA}
  -
  \sum_\alpha U_\alpha|\psi_\alpha)(\psi_\alpha|
  \quad,
\label{eq:hsic}
\end{eqnarray}
\end{subequations}
That form \PGRcomm{embodies the state-dependence in the projector and
displays clearly the non-hermitian nature of $\hat{h}_\mathrm{SIC}$, 
which is also not} invariant under a unitary transform amongst the
$|\psi_\alpha)$.

Variation of the action 
with respect to $\psi_\alpha^*$ thus yields the TDSIC equations as
\begin{subequations}
\label{eq:tdsic}
\begin{eqnarray}
&
  \big(
   {\hat h}_\mathrm{SIC}-\mathrm{i}\hbar\partial_t
  \big) 
  |\psi_{\alpha})
  = 
  \sum_{\beta} |\psi_{\beta})\lambda_{\beta\alpha}
  \quad,
\label{eq:tdsic0a}
\\
  &
  \lambda_{\beta\alpha} 
  = 
  (\psi_{\beta}| \hat{h}_{\alpha} - 
   \mathrm{i}\hbar\partial_t|\psi_{\alpha})
   \quad.
\label{eq:tdsic0b}
\end{eqnarray}
together with the symmetry condition 
\begin{eqnarray}
&  0
  =
  (\psi_\beta|U_\beta-U_\alpha|\psi_\alpha).
\label{eq:symcond_td}
\end{eqnarray}
\end{subequations}
which has to be fulfilled at each instant.  
It should be noted that once one has
achieved the symmetry condition (\ref{eq:symcond_td}), the SIC Hamiltonian
(\ref{eq:hsic}) acquires an interesting property.  Although it remains
non-hermitian as a whole, it becomes hermitian within the space of
occupied states
\begin{equation}
  (\psi_\beta|\hat{h}_\mathrm{SIC}|\psi_\alpha)
  =
  (\psi_\alpha|\hat{h}_\mathrm{SIC}|\psi_\beta)^*
  \quad.
\label{eq:occ-hermitian}
\end{equation}

The above TDSIC equations are quite involved as the time propagation,
\MDcomm{Eqs.~(\ref{eq:tdsic0a}-\ref{eq:tdsic0b})}, is constrained by the
symmetry condition (\ref{eq:symcond_td}),
unlike propagation with a strictly hermitian Hamiltonian (LDA or
Hartree-Fock) for which the symmetry condition is automatically fulfilled.
A simple minded step
\begin{equation}
  |{\psi}_\alpha (t)) =
  \exp{\left\{
    -\frac{\mathrm{i}}{\hbar}\int_{t_0}^{t} \textrm dt'\,
     \hat{h}_{\rm SIC}(t') 
  \right\}} 
  |\psi_\alpha(t_0)) \, ,
\nonumber
\end{equation}
is thus not directly applicable because it 
does not ensure the preservation of the symmetry condition. One 
has to employ an interlaced step which allows to fulfill
simultaneously Eqs.~(\ref{eq:tdsic}).

In order to overcome this difficulty, we note that there is always the
freedom of unitary transformations amongst the set of occupied
orbitals $\{\psi_\alpha,\alpha=1,...,N\}$ without changing the
state of a system. The $\psi_\alpha$ are the unique
ingredients of the SIC mean field. But propagation of the whole state
may be formulated in terms of another set $\{\varphi_i\}$ chosen to have
convenient propagation properties and 
connected to the  $\psi_\alpha$ by a unitary transformation 
within occupied states~:
\begin{subequations}
\label{eq:tdsic-2}
\begin{equation}
  |\varphi_i(t))
  =
  \sum_{\beta=1}^N |\psi_{\beta}(t))\ \upsilon_{i\beta}^*(t)
  \;.
\label{eq:ut-td}
\end{equation}
We call the $\{\varphi_i\}$ the ``propagating'' set and the
$\{\psi_\alpha\}$ the ``symmetrizing'' set, the idea being to perform a
joined propagation of $\{\varphi_i\}$ and $\{\psi_\alpha\}$ such that
each of the two (connected) sets of orbitals contributes either the
propagation or the symmetry condition.  It is nevertheless crucial to
note that the $\psi_\alpha$ remain the key constituents composing the
SIC potentials (\ref{eq:usic}) and entering the symmetry condition
(\ref{eq:symcond_td}) through the $U_\alpha$.
The propagation set  $\varphi_i$ gives
the freedom to choose the propagation within occupied states
at convenience. We choose it such that
\begin{equation}
  \left({\hat h}_{\rm SIC} - \mathrm{i}\hbar\partial_t \right)
  |\varphi_i) 
  = 
  0
  \;,
\label{eq:diag_tdsic}
\end{equation}
which allows to fulfill
\MDcomm{Eqs.~(\ref{eq:tdsic0a}-\ref{eq:tdsic0b})}
and which is possible as soon as the $\psi_\alpha$ fulfill the
symmetry condition (\ref{eq:symcond_td}). 
This allows then to propagate the $\{ \varphi_i \}$  as~:
\begin{equation}
  |\varphi_i(t))
  =
  \exp{\left\{
    -\frac{\mathrm{i}}{\hbar}\int_{t_0}^{t}\, \textrm dt'\,
     \hat{h}_{\rm SIC}(t')\right\}}
  |\varphi_i(t_0)) 
\label{eq:diag_propag}
\end{equation}
and to care for the symmetry condition at the side of the
$\psi_\alpha$, the latter fixing the coefficients of the unitary
transformation. This reads formally
\begin{equation}
  \upsilon_{i\beta}(t)\,:\quad
  0
  =
  (\psi_\beta|
    U_\beta[|\psi_\beta|^2]-U_\alpha[|\psi_\alpha|^2]
  |\psi_\alpha)
  \;.
\label{eq:step2}
\end{equation}
\end{subequations}
It is to be noted that the propagator in Eq.~(\ref{eq:diag_propag}) is
not strictly unitary because $\hat{h}_\mathrm{SIC}$ is not
hermitian. But the hermiticity within occupied space, 
Eq.~(\ref{eq:occ-hermitian}), guarantees that the propagation
(\ref{eq:diag_propag}) preserves orthonormality within occupied
space, {\it i.e.} $(\varphi_i(t)|\varphi_j(t))=\delta_{ij}$.

The above described propagation scheme for TDSIC exhibits some
interesting properties. As already noted, it preserves
orthonormality, which is a crucial requirement for any time evolution
scheme. One can also show that energy \PGRcomm{and the zero-force
theorem are} conserved (as long as
there are no time-dependent external fields). 
Finally, we remark that the choice
(\ref{eq:diag_propag}) for the propagating set is not the only
possibility.  There is some freedom for other choices as, e.g.,
optimizing single-particle energies \cite{Guo07b}.

\begin{figure}[ht]
\centerline{\includegraphics[width=5.5cm,angle=-90]{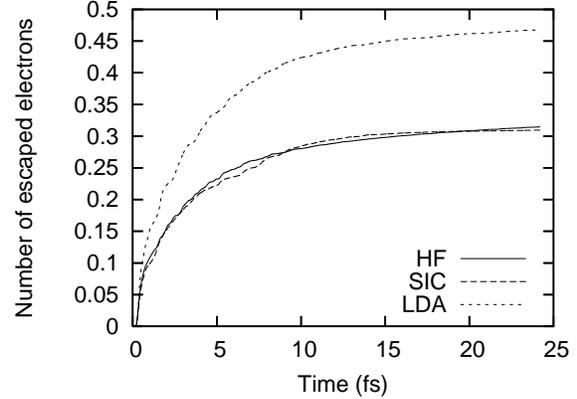}}
\caption{\label{fig:ex} Time evolution of the number of emitted 
electrons for a dimer molecule with two electrons.
Compared are results from full TDHF, TDLDA and TDSIC.
}
\end{figure}
As a final point, we apply TDSIC to a typical example of laser
excitation and subsequent ionization of a molecule. We use a simple
one-dimensional model for a molecule in the spirit of
\cite{Hen98}. The model case is a dimer with two electrons having the
same spin. As an interaction, we use the following smoothed Coulomb
potential, 
$
\displaystyle
  w_{\rm ij} 
  = 
  {e^2}/{\sqrt{(x_{\rm i}-x_{\rm j})^2 + a_{\rm ij}^2}}
$
, where the parameters $a_{\rm ij}$ for electron-electron, electron-ion
and ion-ion interactions are tuned to reproduce typical molecular
energies. Taking that interaction, we develop with LDA an energy
functional for the exchange term. Working at the level of exchange
only allows to have fully fledged time-dependent HF (TDHF)
calculations as benchmark to which DFT calculations can be compared.

A very short laser pulse is simulated as an instantaneous boost
\cite{Cal97b}. This has the advantage that energy conservation can be
used as test for the calculations.  We have checked conservation of
energy, orthonormality and symmetry condition which are all  found  fully
satisfied. As a further observable, we consider the degree of ionization
which, as stated above, is a sensitive quantity to probe the effect of
SIC. 
The results are shown in Fig.~\ref{fig:ex} comparing the TDHF
benchmark with TDLDA and TDSIC. It is obvious that TDSIC comes  very
close to the benchmark.

Finally we want to remark that the double-set strategy can also be
applied to the static SIC problem. 
The static SIC equations are obtained by mi\-ni\-mi\-zing
$E_\mathrm{SIC}$ with a constraint on orthonormality of the
single-particle wavefunctions $\psi_\alpha$, 
$  {\delta}_{\psi^*_\alpha}\left(
   E_\mathrm{SIC} -    \sum_{\beta,\gamma}^{}
   (\psi_{\beta}|\psi_{\gamma})\lambda_{\gamma\beta}
  \right)   =   0$,
{following standard variational derivations} \cite{Ped84,Goe97}.
One then simply obtains
$\hat{h}_\mathrm{SIC}|\psi_\alpha)=
  \sum_{\beta} |\psi_{\beta}) \lambda_{\alpha\beta}$,
again together with the symmetry condition Eq.(\ref{eq:symcond_td}).
Note that the  matrix of Lagrange multipliers
$\lambda_{\beta\alpha}=(\psi_\beta|\hat{h}_\mathrm{SIC}|\psi_\alpha)$
is usually non-diagonal. The states $\psi_{\alpha}$ which
emerge as solutions of stationary equations are
optimized to produce the correct SIC potentials.  They do not give any
clue on single-particle energies.  One can now introduce a second set
of $\{\varphi_i\}$ connected by the stationary analogue of the unitary
transformation (\ref{eq:ut-td}) and require, e.g., that these
$\varphi_i$ diagonalize $\hat{h}_\mathrm{SIC}$ (expressed in terms of 
the $\psi_{\alpha}$, Eq.(\ref{eq:hsic})), or \PGRcomm{equivalently} 
the  matrix of
Lagrange multipliers $\lambda_{\alpha\beta}$.  The
eigenvalues thus obtained can be interpreted as single-particle
energies and they are found to agree fairly well with HF values 
{in the case of exchange only calculations that we consider here}. 
\MDcomm{Note finally that these static SIC equations also emerge naturally as
the stationary limit of TDSIC.}

We have proposed in this letter a consistent variational formulation
of time-dependent SIC (TDSIC) together with a manageable and
transparent scheme for \PGRcomm{the solution} of the TDSIC equations.
This scheme provides conservation of energy, \PGRcomm{zero-force
theorem}, and orthonormality of the occupied states.  The stationary
limit of static SIC is also properly recovered both by the theory
itself and by the propagation scheme. We applied the scheme to
laser-induced ionization of a one-dimensional dimer molecule as a test
case. The calculations have proven to run stable and to fulfill all
theoretical constraints.  As a critical observable, we have
investigated the time evolution of ionization and found nice agreement
of the {exchange only} TDSIC results with an exact TDHF
calculation.

Although even full 3D calculations have proven to be feasible, it is
to be admitted that fully fledged TDSIC is rather involved and thus
computationally complex. We consider it as starting point for further
development towards 
simplified schemes. A promising option is provided by a time-dependent
form of optimized effective potentials \cite{Ull95a,Kue07}.  Work in
that direction is in progress. Nevertheless the full TDSIC serves as a crucial
benchmark for such developments.

\bigskip

\acknowledgments
This work was supported,
by  
%
Agence Nationale de la Recherche (ANR-06-BLAN-0319-02), 
the Deutsche Forschungsgemeinschaft (RE 322/10-1),
and the  Humboldt foundation.

\end{document}